# A Hybrid Cloud ERP Framework For Processing Purchasing Data

Xinyu Zhang

Department of Data Science and Artificial Intelligence, Faculty of Information Technology, Monash University

*Abstract*—Cloud-based enterprise resource planning (cloud ERP) systems have existed in the business market for around ten years. Cloud ERP supports enterprises' daily activities by integrating organizational back-end systems in the cloud environment. One of the critical functions that cloud ERP offers is the purchasing application. The purchasing function of cloud ERP enables enterprises to streamline all the online purchasing transactions in real-time automatically. Even cloud ERP is deployed quite often these days, organizations somehow still lack the knowledge of it; to be specific, there are many issues attached to cloud ERP implementation yet to be solved. Hence, this paper compares four leading cloud ERP platforms in Australia and proposes a hybrid cloud ERP framework to process online purchasing transactions. By adopting a case study approach, a purchasing web-based application is designed and presented in this paper. In general, the proposed hybrid cloud ERP framework and the integrated web-based purchasing application allow user companies to process online purchasing transactions with short operation time and increased business efficiency; in the meantime, the proposed framework also reduces security risks attached to the public cloud.

*Keywords—cloud computing, cloud ERP framework, hybrid cloud, web-based purchasing application*

## I. INTRODUCTION

The term "enterprise resource planning (ERP)" was firstly introduced by the Gartner Group in the early 1990s [1]. ERP system is a software package that helps enterprises deal with their daily business activities by offering them integrated organizational back-end systems, such as purchasing, finance, human resources (HR), sales and inventory, and marketing [3]. With the emergence of cloud computing technology, currently, the leading way of delivering ERP applications is through the cloud environment, and it is called cloud-based enterprise resource planning (Cloud ERP). Cloud computing technology is grabbing increased attention in the business market nowadays; it is considered as the "next-generation platform" to conduct business [47] by providing significant economic benefits for its customers. Cloud computing is a packaged model that offers a set of existing technology bundles for its consumers through the cloud environment [2]. Therefore, cloud ERP is regarded as an innovative approach to conduct business.

Delivering ERP systems in the cloud environment gains greate competitive advantages for its consumers because it offers user enterprises opportunities to focus more on their core business rather than handling massive IT infrastructures by themselves. In the digital economics era of today, increasing numbers of enterprises tend to adopt cloud ERP systems, especially small and medium-sized enterprises (SMEs), due to cost-saving and easy to access characteristics [4].

Even though cloud ERP is still an emerging software application, it will eventually be deployed in a broad range within the business market [5]. A large number of scholars have already identified various factors influencing the adoption decisions of cloud ERP [6, 7, 8], whereas there is a paucity of studies on the implementation of cloud ERP. Therefore, the proposed research questions for this research will be: *What are the benefits of implementing cloud ERP systems, and in what way of the implementation will bring the most significant advantages for the enterprises?*

In order to answer the above research questions, this paper firstly highlights the benefits of implementing cloud ERP systems, then brings a comparison among four leading cloud ERP platforms in Australia to develop a framework for better handling online purchasing processing. Besides, this paper intends to provide an informative knowledge of the cloud ERP implementation, including the opportunities, concerns, as well as challenges. Besides the proposed hybrid cloud ERP framework, a web-based purchasing system for processing purchasing data is also presented in this paper.

The structure of this paper is outlined as follows: Section II presents a comprehensive background of existing cloud ERP relevant literature. Section III indicates the methodology used for this research paper. Section IV demonstrates the comparison among four cloud ERP platforms from four well-known technology companies in Australia. Section V illustrates the proposed hybrid cloud framework of ERP systems for processing online-based purchasing transactions. Section VI adopts a case study approach to present the designed web-based purchasing application for the implementation of the proposed framework. Section VII indicates companies that are suitable to implement the proposed framework. And section VIII concludes the whole paper, along with providing the contributions and future research directions of this paper.

## II. RELEVANT WORK

As is well-known, cloud computing consists of three delivery models: Software-as-a-Service (SaaS), Platform-as-a-Service (PaaS), and Infrastructure-as-a-Service (IaaS); cloud ERP system is delivered through the SaaS model [21]. Besides delivering cloud ERP systems, enterprises can also keep their ERP modules within their infrastructure sitting on the in-house platform, and this way of keeping ERP modules is known as on-premise ERP [22, 30]. In the past few decades, plenty of scholars have focused their attention around on-premise ERP systems, such as their adoption [43, 44], implementation [45], and performance [35, 31]. With the emergence of cloud ERP, in industry, organizations changed the way they used to function; in academia, scholars also focused their attention on the cloud ERP.

Currently, most ERP systems are delivered through the SaaS model in the public cloud, which allows user companies to subscribe to the service provider so that all the add-on modules can be accessed through web browsers without pre-installing any software applications on the user's side [21]. In the past decade, quite a few research studies were conducted around cloud ERP; these relevant studies mainly can be classified into four categories: cloud ERP benefits and drawbacks [21, 22, 25], factors influencing the deployment decisions of cloud ERP [4, 6, 7, 8], migration journey from on-premise ERP to cloud ERP [26, 27], and critical success factors affecting the implementation of cloud ERP [28, 29]. Details on the cloud ERP relevant work and their categories can be viewed in Appendix.

With those four categories, much research was conducted around cloud ERP deployment. Furthermore, many scholars intend to understand the cloud ERP deployment phenomenon from various aspects. For example, most studies adopt the case study approach by conducting interviews with different stakeholders (e.g., client companies, vendors, and consultant companies) to identify cloud ERP adoption factors [30]. Also, much research examines the critical factors influencing cloud ERP adoption decisions made by comparing between SMEs and large-scale enterprises [4]. Furthermore, there are also few studies present cloud ERP adoption factors based on different nations [3, 6]. Generally speaking, these cloud ERP relevant studies conclude that security (i.e., data security, data access authorization, and cloud privacy) [4], accessibility (i.e., whether the system is easily accessed) [21], integration (i.e., system's ability to seamlessly integrate with other systems like legacy systems or any other software applications in the different environment) [22], functionality (i.e., system capabilities) [31], maintainability (i.e., system's backup and maintenance) [24], top management support (i.e., decision-making and support from top managerial level) [6], cost (i.e., operational cost, infrastructure cost, and maintenance cost) [22], and vendor reputation (i.e., vendor's capability and integrity) [23] are the most common factors which have influences on the adoption decisions of cloud ERP.

Besides those identified common adoption factors of cloud ERP, substantial critical success factors that impact the implementation of cloud ERP were also identified by many scholars [28, 29, 36, 37]. In general, data security concerns, customization limitations, and system integration issues are the top three shortages [28] on the implementation of cloud ERP. Among these three concerns, there are always debates about data security issues around cloud ERP applications. Many scholars tend to argue that cloud-based systems expose more security concerns compared to on-premise systems since the user companies do not have full control over the system, which will lead to more uncertainty of accessing authorization, data storage, and network security [22]. Moreover, as for the customization limitations of cloud ERP, since SaaS vendors provide a standardized cloud-based ERP application to all the customers, the customization of cloud ERP is hard to achieve [22]. Additionally, integration between cloud applications and in-house systems is another dilemma that needs to be tackled so that data inconsistency can be avoided [28]. These concerns around cloud ERP are still in controversy, and a specific solution to solve these problems is yet to be given.

Furthermore, it is interesting to see that among all the cloud ERP relevant studies, mainly of these studies were done through the qualitative approach, and these studies dedicated to identify critical factors that have influences on the adoption decision or the implementation performance of cloud ERP. However, studies on investigating cloud ERP sub-modules, or offering solutions to deal with security, customization, and integration issues are lacking. Therefore, this paper tries to add knowledge to the existing literature in terms of cloud ERP implementation and to bring an integrated solution for handling related implementation issues. Therefore, the below section presents the methodology used for this research.

III. METHODOLOGY

This paper intends to bring an integrated framework for the implementation of cloud ERP systems within user enterprises for handling system security, customization, and integration limitations brought by regular cloud ERP. In order to do so, a methodology framework is designed and presented in Figure 1 below.

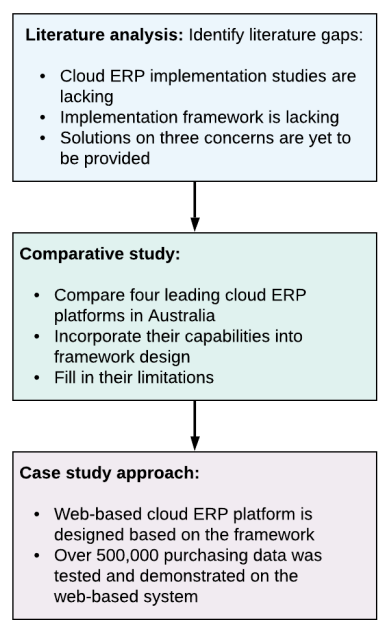

Fig. 1. Methdology approach.

This research was implemented following the above three steps. Firstly, by conducting a literature analysis, research gaps were identified and highlighted in this paper. Since cloud ERP implementation studies are under-researched, plus no ideal solution is suggested for better dealing with cloud system security, customization, and integration issues over cloud ERP implementation. This paper dedicates to understand why it is essential for enterprises to implement cloud ERP systems, and in what kind of way best suits for companies to implement the system.

Secondly, this paper provides a comparative study on four leading cloud ERP platforms in Australia for the development of the framework. By integrating the four platforms' benefits and overcome their limitations, the framework is designed and presented in Section IV. Details on the comparison will be discussed in the next section.

Thirdly, this paper adopts a case study approach for testing the proposed framework. The case study approach is quite useful when trying to understand a phenomenon in great depth [38]. Therefore, the reason for choosing a case study is because we are trying to understand the intentions of cloud ERP implementation for better developing the framework, so that it is efficient to adopt a case study approach to see how the proposed framework performs in a real-life situation. With the proposed framework, a web-based purchasing application is then designed and presented in the paper. By operating the system using the online retail purchasing data retrieved from the UCI machine learning repository [39], 541,909 data were used for operating the system. At last, based on the designed template, the system usability is then discussed. Hence, details on the comparison are presented in Section IV.

## IV. Cloud ERP Platforms Comparison

In spite, there is an existence of studies on identifying factors influencing cloud ERP adoption and implementation, the implementation framework of cloud ERP is still limited. Therefore, to address this issue, this paper compares four cloud ERP platforms from four leading technology companies in Australia. This comparison is used to design a framework for improving the purchasing processing performance for cloud ERP user companies. Below presents the details of the four platforms.

### A. Oracle NetSuite

Oracle ERP Cloud provides enterprises with a set of applications such as finance, accounting, procurement, project management, and risk management to deal with day-to-day organizational activities [9]. Oracle ERP Cloud also offers users with customized dashboards demonstrating real-time transactions, automatically generated reports, user access visibility, and better communications across departments. Furthermore, among all the ERP modules Oracle offers, NetSuite is regarded as the top #1 cloud-based ERP software solution in Australia [10]. NetSuite is mostly suitable for small and medium-sized enterprises (SMEs) to deploy because it is highly scalable and cost-effective [10].

### B. Microsoft Dynamics 365

Microsoft Dynamics 365 is a cloud-based system that includes both ERP applications and customer relationship management (CRM) applications. Microsoft Dynamics 365 is built on the Azure cloud platform, and they aim to seamlessly integrate "database, business intelligence, infrastructure, and ERP together" [11]. Besides, Microsoft Dynamics 365 also offers enterprises various ERP modules like HR, commerce, sales, marketing, finance, operations, and services [12]. By deploying Dynamics 365, enterprises tend to perform more efficiently, operational expenses are reduced, and customer services are also optimized [12].

### C. SAP Cloud ERP

SAP Cloud ERP is a comprehensive application; it not only consists of plenty of business modules but also is compliable and adaptive to changes. Moreover, it has two platforms available for different sizes of organizations (medium size and small size companies) [13]. Besides offering essential functional benefits, SAP Cloud ERP is well-known for its ease of use and secure characteristics [13]. It is regarded as the tool which empowers the next generation for conducting business [13].

### D. MYOB Advanced Cloud ERP

MYOB is namely the number #1 choice of ERP systems in Australia; it offers three types of ERP systems, including cloud-based, on-premise, and ERP for complex business [14]. With MYOB cloud-based ERP system, enterprises can choose different modules like inventory and distribution, financial management, customer management, accounting, and payroll management [15]. MYOB strives to create better value for its customers by providing them professional support, high accessibility, and convenient monthly subscription plans [16].

### E. Comparison Among The Four Platforms

Oracle, Microsoft, SAP, and MYOB are the four leading technology companies in Australia, and they are also the top cloud ERP vendors in Australia. Since these four companies are in significant drives on the cloud ERP implementation in Australia, this paper compares their platforms to develop an integrated framework for better processing online purchasing transactions.

Today, highly competitive pressure exists in the business market for both cloud ERP service providers and enterprises. To be specific, enterprises are struggling to choose the right cloud ERP solution due to the many reasons, such as the price of the system [22], the functionalities that the system offers [31], their business characteristics and requirements [40], as well as various choices of platforms that the vendor offers.

Table 1 below presents the comparison result of these four cloud ERP platforms. This table not only helps for the development of the purchasing framework but also brings some ideas for enterprises to consider when making cloud ERP deployment decisions. And Table 1 compares those platforms from four perspectives: the business modules they offer, enterprise characteristics they suit for, their prices, as well as the limitations they have.

TABLE I. COMPARISON OF THE FOUR LEADING CLOUD ERP PLATFORMS IN AUSTRALIA

| Cloud ERP Platform | Comparison Criteria | | | |
|---|---|---|---|---|
| | *Business Modules or System Capabilities* | *Enterprise Characteristics* | *Cost* | *Limitations* |
| Oracle NetSuite | Finance, Procurement, Risk management, Enterprise performance management (EPM), Project management (PM) | Suitable for SMEs | Low | The platform lacks integration and compliance among modules and systems [17] |
| Microsoft Dynamics 365 | Finance, Marketing, Operations (including Manufacturing and Supply Chain), HR, Sales, Commerce, Service | Suitable for SMEs | Low | The usability and functionality are not as high as other platforms [18] |
| SAP Cloud ERP | Finance, Purchasing, Procurement, Sales, Customer relationship manage, Inventory, Localisation, PM, HR, Analytics reporting, Supply chain (SC) management | Provide platforms for both mediul enterprises and small enterprises | Medium | More costly compared to previous two platforms [19] |
| MYOB Advanced | Finance, Inventory, Payroll management, Distribution, Project Accounting, Customer relationship management | Suitable for all sizes of enterprises | High | More costly compared to all other platforms, limited functionalities and customer support [20] |

As demonstrated in Table 1 above, these four cloud ERP platforms have their unique benefits as well as the limitations. Firstly, compare from the business modules that they offer; notably, all four platforms provide a set of ERP modules in the cloud for their user companies to deal with daily business activities, while financial and purchasing are the two main applications that all platforms offer. It should be noted that these four platforms usually include both purchasing and procurement modules within their purchasing applications. Moreover, the capabilities of their purchasing modules are also varied; for example, the system can modernize warehouse management and streamline supply chain [12], and offer real-time inventory distribution [15]. Each of the purchasing capabilities offered by the four platforms has been added into the framework to guarantee the comprehensiveness of the developed framework capabilities.

Secondly, from the enterprise's characteristic aspect, most of these platforms offer cloud ERP applications to SMEs, while MYOB Advanced is the only platform that is available to both SMEs and large companies. Also, the price of these platforms are varied; due to the high-level system support and customer engagement, as well as the system competencies MYOB offers, MYOB then has the highest price among these four platforms [20].

However, the limitations of these four cloud ERP platforms identified by the user companies also match with the concerns identified from the existing literature [17, 18, 19, 20], as security, integration, and functionality are the three primary issues with implementing cloud ERP [28]. Therefore, to reduce the security concerns with cloud-based systems, to help with system integration limitations, and to offer comprehensive functionalities, a hybrid framework including the integrated purchasing application architecture is presented in the below Section V.

V. PROPOSED HYBRID CLOUD ERP FRAMEWORK

The framework of a hybrid cloud ERP solution with the integrated purchasing application architecture is presented and discussed in this section. The proposed framework intends to accelerate the online purchasing processing, improve the interactive performance between customers and the system, also provide a safe environment for payment services and data storage. And Figure 2 below demonstrates the proposed framework.

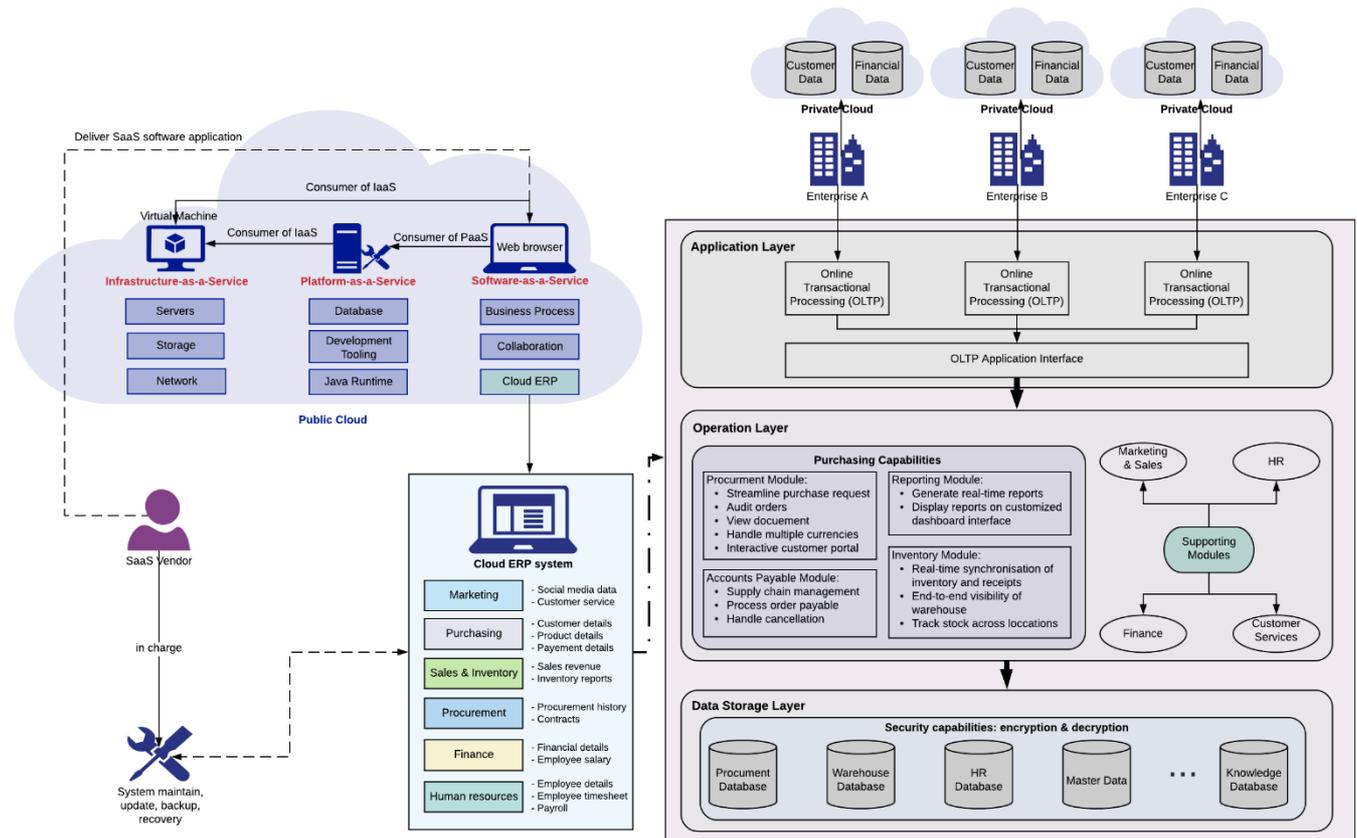

Fig. 2. Proposed Hybrid Cloud ERP Framework

Figure 2 above demonstrates the proposed hybrid cloud ERP framework, including the general cloud ERP delivery cycle, which is displayed on the left-hand side of Figure 2; the designed integrated purchasing application architecture, which is shown on the right-hand side of Figure 2.

On the one hand, the left-side framework (Figure 2) illustrates that the PaaS and IaaS models are both consumers of IaaS [46]. IaaS vendors offer necessary computing and storage capabilities to its consumers, PaaS vendors provide packaged IT capabilities, while SaaS vendors offer software applications using the environment that IaaS and PaaS service providers offer [46]. As cloud ERP is delivered through the SaaS model [21], the SaaS service provider makes the system visible and accessible from web browsers in the public cloud [21] (i.e., public cloud is made available for the public or groups of organizations to access [32]). Therefore, SaaS service vendors can offer different ERP modules to different enterprises through the public cloud environment. In the meantime, SaaS service providers are then responsible for all the system's maintenance, such as the system's upgrades, bug-fixed, backup, and recovery [28]. Moreover, because cloud

ERP systems support various applications for user companies; different organizations have different system's requirements and seek for different ERP modules, such as finance, HR, sales, and marketing; instead of moving all the ERP modules into the cloud environment, the SaaS service provider can simply offer corresponding applications to its user enterprises based on their demands.

On the other hand, the right-hand side of the proposed framework (Figure 2) demonstrates the designed integrated purchasing application architecture of the hybrid cloud ERP solution. Since purchasing application usually includes both purchasing and procurement modules, and they are critical to organizational operations, this study focuses the attention on the performance of online purchasing processing, and the detailed purchasing application is presented using three-layer architecture.

The proposed purchasing application architecture consists of three layers: the application layer, the operation layer, and the data storage layer. Firstly, cloud ERP user companies can access the purchasing system through the online transactional processing (OLTP) application. This application layer handles the actions made by users through the customized interface. Besides, customers' requests will be received as inputs of the transactional process and will be passed to the operational layer, so that after the input is processed, the responses will be sent back to the application layer.

The second layer is the operational layer, which acts as the brain of the entire purchasing system; it processes the inputs made by users. Mainly, the ideal sub-modules supported by the operational layer of this purchasing application are procurement, accounts payable, reporting, and inventory. Moreover, the proposed purchasing application should be able to handle the corresponding capabilities listed in the proposed framework within different sub-modules. For instance, the system will automatically streamline purchase requests, generate real-time reports, support real-time synchronization of inventory and receipts, and demonstrate end-to-end visibility of the warehouse. Notably, the proposed capabilities incorporate the functionalities that the four existing cloud ERP platforms offer from the above four listed companies. Besides the four sub-modules, the operational layer should also be supported by other related modules, such as financial, sales, and customer services. These two parts corporate together to control and provide related data from the storage layer, as well as offering basic functionalities to users. Moreover, those system functions should be offered along with the support of "security, authentication, and visualization control" [33].

The third layer is the data storage layer, where customer requirements are received and stored. Besides, other relevant data which is required for the operation of the system is also stored in this layer, such as purchasing history and product details. Notably, all of these purchasing data is stored in the public cloud environment, and they are managed by the service provider.

Moreover, this proposed framework demonstrates an integrated ERP solution using the hybrid cloud; a hybrid cloud is a combination of using a public cloud and a private cloud [32]. The hybrid cloud solution here keeps most sensitive modules (e.g., financial data and customer data) in the private cloud environment sitting inside organization's infrastructure so that the enterprises have the control over them, while less critical modules (i.e., purchasing history, product details) are kept and running in the public cloud and are managed by the service provider [4]. Using this hybrid cloud approach, it maximizes the user control over the system, and improve data security and information privacy to an extent. Furthermore, data encryption and decryption will still be required to decrease the likelihood of data leakage and third-party concerns. Meanwhile, the performance of the proposed purchasing processing application will also be improved due to the comprehensive functionalities offered.

In general, with the innovation of technology, cloud-based ERP systems play a significant role in the business market for enterprises to achieve innovation, competitive advantage, and organizational sustainability.

Despite the existence of cloud ERP studies from the academic perspective, the industry-related issues around cloud ERP are hardly resolved; the implementation of cloud ERP is minimal, not even mention the implementation framework. Therefore, the proposed framework intends to help client companies understand the delivery cycle of cloud ERP, offer sufficient purchasing capabilities to increases the performance of online purchasing processing, as well as reduces security concerns raised before implementing the system. Additionally, using the case study approach, the proposed hybrid framework will be operated to test its usability by applying a designed web-based purchasing application; therefore, situations like enterprises do not quite satisfy with the performance of the system after implementing specific cloud ERP modules, will be well avoided. And the case study implementation is discussed in Section VI below.

## VI. CASE STUDY

As demonstrated in the methodology framework (i.e., Fig. 1), a case study approach is utilized in this research study for implementing the proposed hybrid cloud ERP framework. Based on the proposed framework, a web-based purchasing application web system is designed; using the 541,909 online purchasing retail datasets retrieved from the UCI Machine Learning Repository [39], the web-based system is operated, and the system template is shown below.

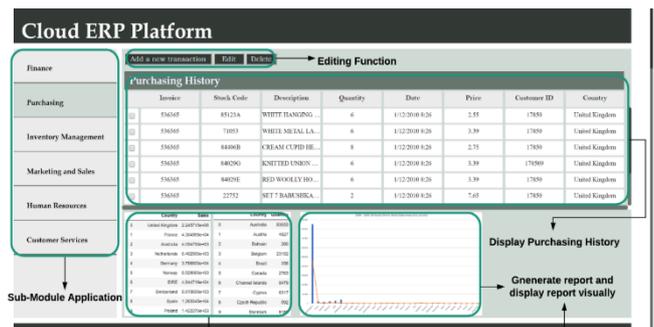

Fig. 3.  Web-based Cloud ERP Purchasing Application.

The purchasing application architecture is suggested to be designed as the above-displayed web system shown in Figure 3. Figure 3 mainly presents two sections: the left-hand toggle bar and the right-hand purchasing application capabilities. First of all, the toggle bar on the left-hand side is used for user companies to access any sub-modules of cloud ERP systems that they subscribe; for example, they can access financial, HR, and other applications by simply clicking the corresponding section.

Additionally, the purchasing capabilities are presented on the right-hand side. In general, this web-based purchasing application is designed based on the proposed hybrid framework; therefore, all the capabilities mentioned in the framework are suggested to be developed in this web-based system, yet only a few functionalities are presented using the template above due to the demonstration limitations.

Quite a few main built-in capabilities are demonstrated in this template; the purchasing history is available to view, edit, and delete on the main page of this web-based system, and the purchasing history is demonstrated based on the latest transaction. Moreover, the full history of the purchasing can be viewed using the right scroll bar; the 541,909 online retail data retrieved from the UCI Machine Learning Repository is listed in this section. Additionally, the below two images present the reporting function: generated real-time purchasing reports based on the purchasing history. Furthermore, the system also can automatically present purchasing details visually using graphic charts.

The proposed hybrid cloud ERP framework, as well as the designed purchasing application template, provide companies with higher chances of improving their business performance by adopting technology innovations. With a large amount of data running in the designed web-based system, the system's operation is very promising; hence, by implementing the proposed framework, two major benefits can be achieved. The proposed framework can help user companies to achieve the system's scalability and increase business efficiency. Therefore, for better providing client companies with guidance on the framework implementation, the next section presents a discussion in terms of the suitable companies to implement the proposed framework and the designed template.

## VII. Suitable Companies To Implement The Proposed Hybrid Cloud ERP Framework

The proposed framework brings a hybrid solution for enterprises to access ERP functionalities in order to process online purchasing transactions. This proposed framework is especially suitable for medium-sized and large-scale companies. According to Australian Bureau of Statistics, a small company is defined as a company that has less than 20 employees; a medium business is defined as a company has more than 20 but less than 200 employees; a large company is defined as an organization has more than 200 employees [34].

For small-size organizations, the most effective approach to manage business activities is to implement full ERP modules in the public cloud environment managed by the service provider. Because managing an IT infrastructure is very costly from both the financial aspect and the human resources aspect, small companies usually cannot invest much in the IT upfront infrastructure expense, operation cost, and maintenance cost [22]. Therefore, by implementing a hybrid cloud ERP solution, expenses over the system's maintenance and resources will not be reduced. Based on this cost issue, this hybrid solution is not very suitable for a small company, particularly not recommendable for startups.

As for medium and large size organizations, this proposed framework is highly recommended. Three major benefits will be achieved by implementing this proposed hybrid cloud ERP framework, and they are: increase system scalability, increase business efficiency, and reduce security concerns.

**Increase system scalability and flexibility.** One of the major benefits of implementing the proposed hybrid cloud ERP framework is to increase system scalability and flexibility; this statement is supported by many scholars [41, 42, 23, 25]. By implementing the proposed framework and the purchasing application in the public cloud, the SaaS service provider will offer scaled resources for the user enterprise based on their requirements and needs [23]. Expressly, different sizes of organizations will be provided with dynamic resources; this is also the reason why this hybrid solution is suitable for both medium enterprises and large enterprises. Moreover, whether the purchasing transactions are big or small, the system's performance will be guaranteed by the provider due to the high scalability of cloud-based systems.

**Increase business efficiency.** Implementing the proposed hybrid cloud ERP framework will highly increase the user company's business performance. The purchasing module, one of the main functionalities that business offers, is undoubtedly an essential aspect for conducting business. From the proposed framework and the demonstrated web-based system, besides the purchasing application supports all the capabilities indicated from the proposed framework, the purchasing transactional process time will be dramatically decreased compared to the traditional way of conducting business manually. By using the online retail purchasing history, the system is performed quite well with short running time and a fast response rate [4, 21, 25]. Managing a large amount of transaction data in the public cloud by the vendor will highly reduce the pressure that enterprises are facing, comparing to managing the data by themselves, which is time-consuming and labour resources-consuming. Moreover, applying the proposed framework and deliver business in this way, the user enterprises get to focus on their core competencies so that the business efficiency will be increased [22], this, in turn, will contribution to creating customer engagement.

**Increase data security and privacy.** In addition to the above two benefits, another benefit is the increased data security and information privacy. To be specific, the main benefit of implementing a hybrid cloud solution rather than a full public cloud is increased data security. Unlike managing data in the private cloud environment, the public cloud exposes more potential security concerns since the data is not kept on the user's side, and the users do not have control over the data [25]. Therefore, keeping sensitive data such as financial data and master customer data in the private cloud managing by the enterprise itself is an appropriate and reliable way.

Nevertheless, security concerns can be regarded as both technical concerns and human-related concerns, even if the technical concerns are reduced, the human-related security issues like data leakage cannot be fool-proofed; therefore, this proposed framework only ensures data security, authenticity, and reliability to some extent. Hence, the access authorization, encryption, and decryption of data within the private cloud environment should still be maintained. And the next section concludes the paper, highlights the contribution, as well as presents the limitations and future research directions.

## VIII. Conclusion, Contribution, and Limitation

To conclude, due to the innovative cloud computing technology, cloud-based ERP systems are now in a significant drive of conducting business in the market. Due to the limitation of cloud ERP implementation studies, this paper contributes both theoretically and practically.

Firstly, in a theoretical way, this research paper adds some knowledge to the existing cloud ERP relevant studies by highlighting the importance of deploying cloud ERP systems as well as the implementation issues. By identifying that security, customization, and integration are the three major concerns attached to cloud ERP implementation, a hybrid cloud ERP framework is developed. The proposed framework demonstrates the delivery cycle of cloud ERP systems and the integrated purchasing architecture of cloud ERP. This framework can then be utilized for a better understanding of cloud computing's structure, as well as understanding cloud ERP implementation and solve corresponding issues.

Secondly, from a practical perspective, this paper helps client companies to understand cloud ERP, as well as helping them to make decisions on the choice of selecting the right system. Additionally, by applying a case study approach with a set of online retail purchasing data, a designed purchasing web-based system is presented in this paper. This web-based system can then be utilized for vendors to deliver hybrid cloud ERP solutions for their user companies.

In general, three major benefits can be achieved by implementing the proposed hybrid framework and the designed template: increase system scalability, increase business performance, and reduce security concerns attached to the public cloud. Furthermore, by implementing the proposed framework, organizations gain higher chances of achieving their business goals.

Nevertheless, even this paper proposes a comprehensive hybrid cloud ERP framework, evaluations of the framework's performance is lacking due to time limitations. Therefore, for future research, the proposed framework and the web-based system should be evaluated. By applying a simulation testing, the 541,909 datasets should be operated with the developed web-based system so that the system performance can be evaluated (i.e., the system's running time and capabilities). Additionally, the performance of the proposed framework and the web-based system should be compared to the existing cloud ERP platform; this can then contribute to the framework re-construction.

APPENDIX

| Category | Literature | Author |
|---|---|---|
| Benefits & Drawbacks | ERP in the Cloud–Benefits and Challenges | Lenart |
| | Potential concerns and common benefits of cloud-based enterprise resource planning (ERP) | Parthasarathy |
| | An analysis of the perceived benefits and drawbacks of cloud ERP systems: A South African study | Scholtz & Atukwas |
| | Benefits and challenges of cloud ERP systems–A systematic literature | Elmonem et al. |
| Migration Journey | Moving from evaluation to trial: How do SMEs start adopting Cloud ERP? | Salim et al. |
| | Switching toward Cloud ERP: A research model to explain intentions | Mezghani |
| | Understanding Intentions to Switch Toward Cloud Computing at Firms' Level: A Multiple Case Study in Tunisia | Hachicha & Mezghani |
| Adoption Factors | Factors affecting cloud ERP adoption in Saudi Arabia: An empirical study | AlBar & Rakibul |
| | Factors affecting the adoption of enterprise resource planning (ERP) on cloud among small and medium enterprises (SMES) in Penang Malaysia | SMALL |
| | Competition and challenge on adopting cloud ERP | Weng & Hung |
| | Technological, organizational, and environmental factors affecting the adoption of cloud enterprise resource planning (ERP) systems | Kinuthia |
| | Exploring determinants of cloud-based enterprise resource planning (ERP) selection and adoption: A qualitative study in the Indian education sector | Das & Dayal |
| | Factors for adopting ERP as SaaS amongst SMEs: The customers vs. Vendor point of view | Rodrigues et al. |
| | Cloud and traditional ERP systems in small and medium enterprises | Saini et al. |
| | Factors that determine the adoption of cloud ERP: A global perspective | Arinze & Anandarajan |
| | Using the Multi-Theory Approach to Investigate the Factors that Affect the Adoption of Cloud Enterprise Resource Planning Systems by Micro, Small and Medium Enterprises in the Philippines | Caguiat et al. |
| | Cloud ERP Adoption Opportunities and Concerns: A Comparison between SMEs and Large Companies | Johansson et al. |
| | Cloud ERP: a new dilemma to modern organisations? | Peng & Gala |
| | Motives and Barriers to Cloud ERP Selection for SMEs: A Survey of Value Added Resellers (VAR) Perspectives | Garverick |
| | Indian SMEs Perspective for election of ERP in Cloud | Mahara |
| | Cloud-based ERP solution for modern education in Vietnam | Nguyen et al. |
| | Cloud ERP Adoption - A Process View Approach | Salim |
| | In-house versus in-cloud ERP systems: a comparative study | Elragal et al. |
| | ERP System Adoption: Traditional ERP Systems vs. Cloud-Based ERP Systems | Al-Ghofaili & Al-Mashari |
| Critical Success Factors of Implementation | Identification of challenges and their ranking in the implementation of cloud ERP: A comparative study for SMEs and large organizations | Gupta et al. |
| | Organizational, technological and extrinsic factors in the implementation of cloud ERP in SMEs | Gupta et al. |
| | ERP on Cloud: Implementation strategies and challenges | Appandairajan et al. |
| | Critical success factors model for business intelligent over ERP cloud | Emam & Ahmed |
| | Cloud ERP implementation | Carutasu et al. |
| | Compliance, network, security and the people related factors in cloud ERP implementation | Gupta et al. |
| Other | Role of cloud ERP on the performance of an organization: contingent resource-based view perspective | Gupta et al. |
| | A framework for evaluating cloud enterprise resource planning (ERP) systems | Chandrakumar et al. |
| | Academic Cloud ERP Quality Assessment Model | Surendro & Olivia |